\documentclass[10pt,conference]{IEEEtran}
\IEEEoverridecommandlockouts
\usepackage{cite}
\usepackage{amsmath,amssymb,amsfonts}
\usepackage{algorithmic}
\usepackage{graphicx}
\usepackage{textcomp}
\usepackage{xcolor}
\def\BibTeX{{\rm B\kern-.05em{\sc i\kern-.025em b}\kern-.08em
    T\kern-.1667em\lower.7ex\hbox{E}\kern-.125emX}}

\usepackage{amsmath}
\usepackage{amssymb}
\usepackage{bm,braket}
\usepackage{float} 
\usepackage{tikz} 
\usetikzlibrary{quantikz}

\usepackage{hyperref}
\usepackage{multirow}
\hypersetup{
     colorlinks   = true,
     citecolor    = blue
}
\def\be{\begin{equation}}
\def\ee{\end{equation}}
\def\bea{\begin{eqnarray}}
\def\eea{\end{eqnarray}}
\def\bes{\begin{eqnarray}}
\def\ees{\end{eqnarray}}
\def\bi{\begin{itemize}}
\def\ei{\end{itemize}} 

\usepackage{amsthm}

\theoremstyle{definition}

\usepackage{tikz}
\usetikzlibrary{braids}

\begin{document}
\title{Post-Variational Ground State Estimation via QPE-Based Quantum Imaginary Time Evolution  } 

\author{\IEEEauthorblockN{1\textsuperscript{st} Nora Bauer}
\IEEEauthorblockA{\textit{Department of Physics and Astronomy} \\
\textit{The University of Tennessee}\\
Knoxville, USA \\
nbauer1@vols.utk.edu}
\and
\IEEEauthorblockN{2\textsuperscript{nd} George Siopsis}
\IEEEauthorblockA{\textit{Department of Physics and Astronomy} \\
\textit{The University of Tennessee}\\
Knoxville, USA \\
siopsis@tennessee.edu}}
\maketitle 
\begin{abstract}
Quantum phase estimation (QPE) plays a pivotal role in many quantum algorithms, offering provable speedups in applications such as Shor's factoring algorithm. While fault-tolerant quantum algorithms for combinatorial and Hamiltonian optimization often integrate QPE with variational protocols—like the quantum approximate optimization Ansatz or variational quantum eigensolver—these approaches typically rely on heuristic techniques requiring parameter optimization.
In this work, we present the QPE-based quantum imaginary time evolution (QPE-QITE) algorithm, designed for post-variational ground state estimation on fault-tolerant quantum computers. Unlike variational methods, QPE-QITE employs additional ancillae to project the quantum register into low-energy eigenstates, eliminating the need for parameter optimization.
We demonstrate the capabilities of QPE-QITE by applying it to the low-autocorrelation binary sequences (LABS) problem, which is a higher order optimization problem that has been studied in the context of quantum scaling advantage. 
Scaling estimates for magic state requirements are provided to assess the feasibility of addressing these problems on near-term fault-tolerant devices, establishing a benchmark for quantum advantage. Moreover, we discuss potential implementations of QPE-QITE on existing quantum hardware as a precursor to fault tolerance. 
\end{abstract}


\section{Introduction} 
As the quantum community approaches to the era of early-fault-tolerant quantum hardware, the demand for practical quantum algorithms that can demonstrate quantum advantage on real-world applications is rapidly increasing. Many quantum algorithms offer exponential theoretical advantage, such as Shor's factoring algorithm \cite{Shor_1997}, but are restricted to fault-tolerant devices. There is also evidence for heuristic algorithms, such as the quantum approximate optimization algorithm (QAOA), to obtain scaling advantage on problems such as combinatorial optimization \cite{Shaydulin_2024}. However, observing quantum advantage will require both near-term fault-tolerant implementations of algorithms, such as transpiling to basis sets used by developing hardware, and establishing problem and algorithm pairings which have the potential for verifiable quantum advantage. 

Recent efforts have been devoted to establishing thresholds for quantum advantage on near-term fault-tolerant hardware, for example, with QAOA (quantum approximate optimization algorithm) solving the Max8SAT combinatorial optimization problem \cite{omanakuttan2025thresholdfaulttolerantquantumadvantage}, where the runtime is compared with state of the art classical solvers. Additionally, fault tolerant algorithms for computing ground state energy estimations have been formulated for quantum problems, such as quantum spin models, which offer a different path toward quantum advantage when compared with classical methods \cite{Kiss2025earlyfaulttolerant,Dong_2022}. Additionally, since near-term fault tolerant devices will likely still have a finite error rate, the scaling of resource requirements with precision or probability of success is pertinent to near-term advantage \cite{Wang_2023,Wang_2025,Lin_2020,Lin_2022,Zhang_2022}.

Quantum Phase Estimation (QPE) is a key component to many quantum algorithms and has been studied for near-term fault-tolerant devices \cite{Goto_2014}. QPE is powerful for spectral calculations in that, given some initial state, it can prepare both the eigenstate of a Hamiltonian and its corresponding eigenvalue encoded in a register \cite{kitaev1995}. However, the probability of obtaining any eigenvalue is proportional to the initial state's overlap with that eigenvector, so most applications for Hamiltonian optimization require pairing with a variational algorithm for initial state preparation. These variational algorithms require parameter optimization, and many forms are known to suffer from the vanishing gradient variance, making optimization ineffective at large scales \cite{McClean_2018}. To address this, recent work has proposed theoretical bounds on the overlap of arbitrary initial states with eigenstates, thereby providing performance guarantees for QPE even with suboptimal inputs \cite{lin2025boundswavefunctionoverlaphamiltonian}. These insights motivate algorithmic alternatives to variational state preparation.

In this work, we develop a quantum algorithm suitable for Hamiltonian minimization and combinatorial optimization on near-term fault tolerant quantum hardware based on QPE, but with a non-variational Quantum Imaginary Time Evolution (QITE) step that projects into the low energy states. The QITE algorithm has emerged as a promising quantum algorithm and has been implemented to various use cases ranging from ground state energy calculations of molecules \cite{Motta_2019} to solving optimization problems \cite{alam2023solvingmaxcutquantumimaginary,bauer2023combinatorialoptimizationquantumimaginary}, and is based on the concept of infinite cooling of a quantum state. In general, the QITE operation is non-unitary, which we implement here by projecting an ancillary qubit entangled with the QPE register into a specific basis state. We study the example of the Low Autocorrelation Binary Sequences (LABS) combinatorial optimization problem, and show that QPE-QITE successfully projects into the ground state for reasonably large imaginary time $\tau$. For the implementation on fault tolerant quantum devices, we transpile the QITE operator into the Clifford + Toffoli (T-gate) basis set and provide estimates for T-gate requirements and scaling. Finally, we provide insights as to how QPE-QITE could be implemented before fault tolerance on current quantum devices. 

Our work is organized as follows. In Section \ref{sec:method}, we introduce the standard QPE method and extend it to QPE-QITE, and also introduce the LABS problem which we take as a case study. In Section \ref{sec:results}, we show results for QPE-QITE on the LABS problem and provide a resource estimation for T-gate requirements on near-term fault-tolerant devices. Finally, in Section \ref{sec:conclusion}, we summarize our results and offer directions for QPE-QITE implementation before fault-tolerance. 













\section{Method}\label{sec:method}  
Here we detail the QPE-QITE method. We will begin with a review of the QPE algorithm, and then build on it to perform QITE. QPE-QITE uses three sets of qubits: the state qubits $S$ which encode the problem (i.e. the Hamiltonian to minimize), the register qubits $R$ which are standard to QPE and encode the estimated energy of the state, and the QITE ancillary qubit $A$ which is used to project the state and corresponding energies to low energies, thus minimizing the Hamiltonian. We also introduce the LABS problem which we wish to study. 
\begin{figure*}[ht!]
    \centering
    \includegraphics[width=0.95\linewidth]{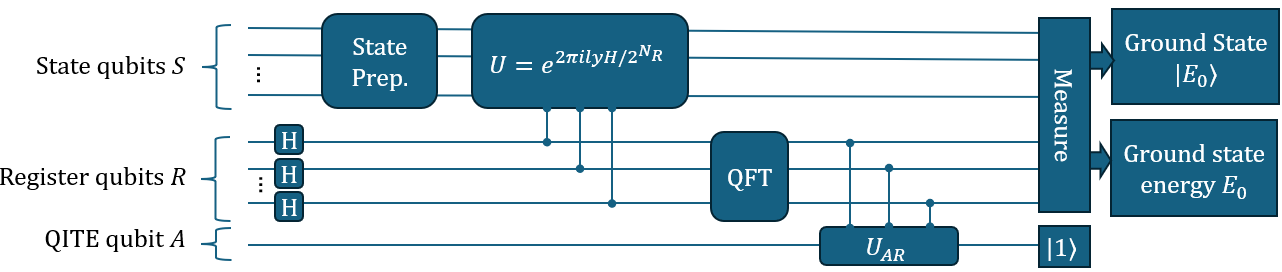}
    \caption{Depiction of the QPE-QITE algorithm as a quantum circuit. }
    \label{fig:qpeqite_drawing}
\end{figure*}
\subsection{QPE Method }
Standard QPE is defined in the context of a Hamiltonian eigenvalue problem,
\be H \ket{E_i} = E_i \ket{E_i} \, \ \ee
where we are interested in the ground state $\ket{E_0}$ with energy $E_0$. 
QPE starts with the state qubits in some initial state $\ket{b}_S = \sum_i b_i \ket{E_i}_A$, where it is required that $b_0\ne 0$. Then, the register $R$ of $N_R$ qubits in the unnormalized state 
\be \ket{s}_R \propto \sum_{y=0}^{2^{N_R}-1} \ket{y}_R\ee 
is prepared and entangled with the state qubits by applying the unitary
\be U = e^{2\pi i l yH/2^{N_R}} \label{eq:ueq}~, \ee
where $l$ and $N_R$ can be adjusted at will. They can be adjusted so as to zoom into the vicinity of the ground state energy $E_0$. We obtain the state
\be U \ket{b}_S \ket{s}_R \propto \sum_i \sum_{y=0}^{2^{N_R}-1} b_ie^{2\pi i l yE_i/2^{N_R}}\ket{E_i}_S\ket{y}_R ~. \ee
Next, the register is Fourier transformed by applying the unitary $U_{QFT}$, where
\be U_{QFT} \ket{y} \propto \sum_{p=0}^{2^{N_R}-1} e^{-2\pi i py/2^{N_R} } \ket{p} ~. \ee
We obtain
\be U_{QFT}U \ket{b}_S \ket{s}_R \propto \sum_i \sum_{p=0}^{2^{N_R}-1} a_{i,p} b_i\ket{E_i}_S\ket{p}_R ~, \ee
where
\be a_{i,p} = \sum_{y=0}^{2^{N_R}-1} e^{2\pi i (l E_i -p)y/2^{N_R}} = \frac{e^{2\pi i l E_i } -1}{e^{2\pi i (l E_i -p)/2^{N_R}} -1}~.\ee
A measurement of the register yields the probability distribution
\be \mathcal{P} (p) \propto \sum_i |a_{i,p} b_i|^2  = \sum_i |b_i|^2 \frac{\sin^2 \pi lE_i}{\sin^2 \frac{\pi (lE_i -p)}{2^{N_R}}}~,\ee
This probability distribution has peaks at $p \approx lE_i$, providing estimates of the energy levels $E_i \approx \frac{p}{l}$, where $p = 0,1,\dots, 2^{N_R}-1$. We want to choose $N_R$ and $l$ so that we capture the peak at $E_0$ and with good enough resolution so that it is resolved from the first excited state peak.

\subsection{QITE} 

QPE-QITE builds upon the QPE algorithm, where before we measure the register, we perform QITE on the register to project to the low energy states. To implement QITE, we attach an ancillary qubit in the state $\ket{0}_A$ and entangle it with the rest by applying the controlled-QITE unitary
\be U_{AR} = e^{i e^{-p\tau} Y_A} \label{eq:uar} ~, \ee
which rotates the ancilla by an angle controlled by the register. We obtain the state
\bea \ket{\Psi} &\propto& U_{AR} U_{QFT}U \ket{b}_S \ket{s}_R \ket{0}_A \nonumber\\ &\propto& \sum_{i,p}  a_{i,p} b_i \ket{E_i}_S \ket{p}_R \left( \cos e^{-p\tau} \ket{0}_A + \sin e^{-p\tau} \ket{1}_A \right) ~. \nonumber\\ \eea
For large $\tau>0$, we have $\sin e^{-p\tau} \approx e^{-p\tau}$. By measuring the ancilla and projecting it onto $\ket{1}_A$, we obtain the state
\be {}_A \braket{1 | \Psi} \propto \sum_{i,p}  e^{-p\tau} a_{i,p} b_i \ket{E_i}_S \ket{p}_R ~,  \label{eq:proj} \ee
implementing QITE.
As $\tau\rightarrow\infty$, $e^{-p\tau}\ket{p}_R\rightarrow\ket{p=0}_R$, 
\be {}_A \braket{1 | \Psi} \propto \sum_{i}  a_{i,p=0} b_i \ket{E_i} \ket{p=0}_R   \propto b_0 \ket{E_0} \ket{p=0}_R~, \ee
where the quantum ground state $\ket{E_0}$ is prepared for $b_i\neq 0$ and the zero energy register state $p=0$ is prepared. 

\subsection{LABS Problem}
The goal of the LABS problem is to minimize the ``sidelobe  energy'' for a system of $N$ spins $\sigma_i\in\{+1,-1\}$ with autocorrelation $\mathcal{A}_k(\bm{\sigma})$ : 
\be \mathcal{E}_{\mathrm{sidelobe}}(\bm{\sigma})=\sum_{k=1}^{N-1}\mathcal{A}_k^2(\bm{\sigma}),~\mathcal{A}_k(\bm{\sigma})=\sum_{k=1}^{N-k} \sigma_i \sigma_{i+k}~.\ee 
This can be mapped to the ground state problem of the quantum Hamiltonian \cite{Shaydulin_2024} 
\bea  \bm{\mathcal{H}}^{\mathrm{LABS}} &=& 2\sum_{i=1}^{N-3}\sum_{t=1}^{\frac{N-i-1}{2}}\sum_{k=1}^{N-i-t}Z_iZ_{i+t}Z_{i+k}Z_{i+t+k} \nonumber\\
&& +\sum_{i=1}^{N-2}\sum_{k=1}^{\frac{N-i}{2}}Z_iZ_{i+2k}~.\eea
The quality of a solution can be quantified by the overlap with the ground states, $P(\bm{\sigma})$. 

For testing QPE-QITE on the LABS problem, the Hamiltonian $\bm{\mathcal{H}}^{\mathrm{LABS}}$ is offset by some constant term $-\alpha \mathbb{I}$, which is ideally the ground state energy. In practice, $\alpha$ is a reasonable estimate for the ground state energy (e.g., from a classical heuristic algorithm) which is then decreased as far as successful. Additionally, since the LABS ground state will be a computational basis state, we choose the initial state as the equal superposition of all computational basis states $\ket{b}_S=\ket{+}^N$. 

\section{Results}\label{sec:results} 
We first test the performance and scalability of QPE-QITE by computing the minimum imaginary time parameter $\tau$ that can prepare the LABS problem solution with $\geq 0.999$ probability after projection. Additionally, we calculate the probability of projecting into the QITE state (where the ancilla is in the $\ket{1}_A$ state), which is the probability of success. 

\begin{figure}[ht!]
    \centering
    \includegraphics[width=1.0\linewidth]{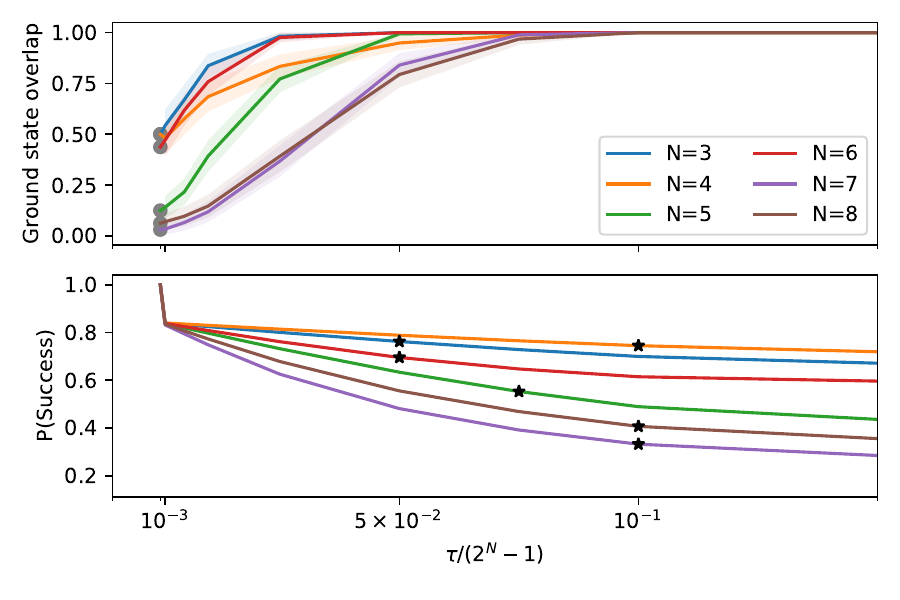}
    \caption{QPE-QITE performance on small LABS problems of size $N$. The upper plot gives the ground state overlap after ancilla post-selection as a function of the imaginary time parameter $\tau/(2^N-1)$. The circles give the probability of measuring the ground state without performing QITE, with only QPE applied to the equal superposition $\ket{+}^N$. The lower plot gives the probability of projecting the ancilla into the $\ket{1}_A$ state as a function of $\tau/(2^N-1)$. The stars give the minimum $\tau/(2^N-1)$ for which the ground state overlap is $\geq 0.999$. } 
    \label{fig:qpeqite_labs_small}
\end{figure}

These results for the LABS problem are given in Figure \ref{fig:qpeqite_labs_small}. Problems up to $N=8$ are studied, where up to $5$ register qubits and 1 ancillary QITE qubit are used. The ground state overlap after projection of the QITE qubit into the $\ket{1}_A$ state is given as a function of $\tau/(2^N-1)$, where modest $\tau$ values are required to obtain convergence to the ground state. The gray dots represent the ground state overlap using only QPE without QITE, where the overlap is the probability of measuring the ground state in the equal superposition $\ket{+}^N$. Note that the LABS ground state is generally degenerate. The probability $P($success$)$ is also shown, which is the probability that the QITE qubit is projected into the $\ket{1}_A$ state. The stars here indicate the minimum $\tau/(2^N-1)$ for which the produced state is converged to the ground state, indicated by a ground state overlap $>0.999$. 

Although QPE-QITE can prepare the ground state for large $\tau\rightarrow\infty$, the practicality of implementing the $U_{AR}$ operator and the gate budget on quantum hardware is a limiting factor in the performance on near-term fault tolerant devices. In order to quantify this, we can study how the operator $U_{AR}$ scales in difficulty with increasing $\tau$. 

For the fault tolerant quantum computer resource estimates, we will use the Clifford $+$ T-gate library, as is standard. Thus, the resource estimation will focus on the number of T-gates (or "magic states") required to perform the algorithm. Note that this count includes both the T-gates and the inverse T-dagger gates. The T-gate scaling of the $U$ operator (for combinatorial optimization problems such as LABS) is 
\be\mathcal{O}(N_R\log^{3.97}(1/\epsilon)|H|)\ee 
where $|H|$ is the number of terms in the Hamiltonian. The $N_R$-qubit QFT T-gate scaling is well-known in literature to be $\mathcal{O}(N_R^2 \log N_R)$, with the approximate QFT further reducing the gate count to $\mathcal{O}(N_R \log^2 N_R)$ with some additional dependence on approximation error $\epsilon$ \cite{park2024tcountoptimizationapproximatequantum}. Now we turn to the last component of the QPE-QITE algorithm, the T-gate scaling of the controlled-QITE operator $U_{AB}$. 

For approximating $U_{AB}$, we can use the Solovay-Kitaev algorithm \cite{Kitaev_1997,dawson2005solovaykitaevalgorithm} to discretize into the Clifford $+$ T-gate basis and compute the error of the decomposition. 
We can expand the second exponential in \eqref{eq:uar} in a Taylor series to calculate the form of $U_{AR}$ for growing $\tau$: 
\be U_{AR} \approx \exp\left[i \left(1-p\tau+\frac{p^2\tau^2}{2}+\mathcal{O}(\tau^3) \right) Y_A\right] ~.\ee
In the following, we obtain the terms by order and their corresponding T-gate budgets: 
\begin{enumerate}
    \item Order $0$: $Y$ rotation on the ancilla, with a fixed complexity of $\mathcal{O}(\log^{3.97}(1/\epsilon))$, where $\epsilon$ is the maximum error per single qubit rotation. 
    \item Order $1$: controlled rotation of the form \eqref{eq:ueq} with Hamiltonian $Y_A$, which requires $2N_R$ CNOT gates and $3N_R$ single-qubit rotations (Y or Z axis), which adds a $\mathcal{O}(N_R\log^{3.97}(1/\epsilon))$ Toffoli complexity, where $\epsilon$ is the maximum error per single qubit rotation. 
    \item Order $2+$: Higher-order terms in the Taylor expansion involve increasingly complex multi-controlled operations that do not admit simple closed-form decompositions. As such, we defer to numerical synthesis and resource estimation for these contributions, using Solovay-Kitaev decomposition to evaluate their impact on circuit depth and T-gate cost. 
\end{enumerate}
Note that these results are an upper bound for large $\tau$, where we will leave more realistic calculations to further numerical experiments. 

\begin{figure}[ht!]
    \centering
    \includegraphics[width=0.95\linewidth]{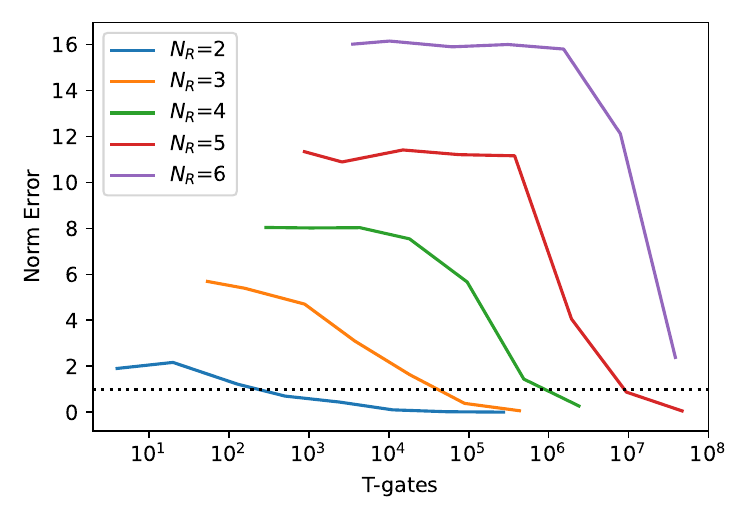}
    \caption{Error vs.\ number of T-gates for the $U_{AR}$ operator transpiled to the Clifford + T-gate basis set for various register sizes. } 
    \label{fig:tgate}
\end{figure}

T-gate numerical calculations for the full $U_{AR}$ operator implemented on a fault tolerant device are given in Figure \ref{fig:tgate}. Here, we compute the $U_{AR}$ circuit representation for large $\tau$ and transpile it to the fault-tolerant Clifford+T-gate library using the Solovay-Kitaev algorithm. The recursion depth of the algorithm is increased to improve the norm error, generating the various points with different T-gate counts. A general threshold of a norm error of $1$ is plotted in the dotted line.  Note that these results are for large $\tau$, where in practice smaller $\tau$ values will allow for more small angle rotations and higher order terms to be removed, reducing the T-gate count. 

\begin{figure}[ht!]
    \centering
    \includegraphics[width=0.95\linewidth]{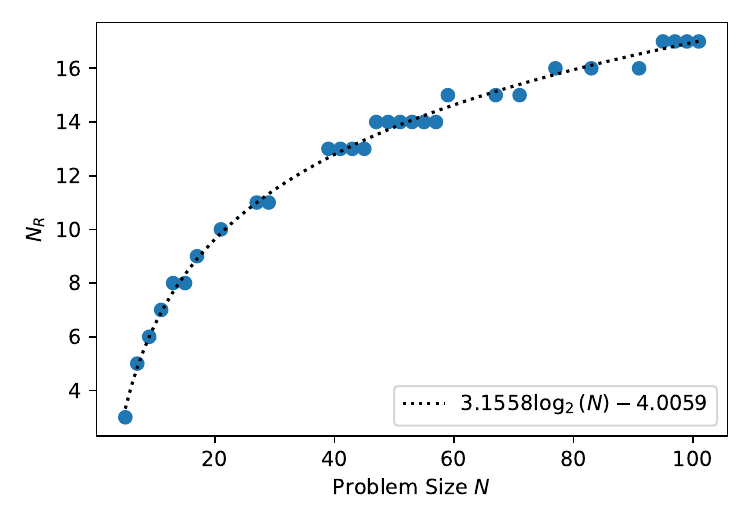}
    \caption{Number of register qubits $N_R$ required to resolve the ground state energy for LABS of problem size $N$. The black dotted line denotes a logarithmic fit with equation form given in the legend. }
    \label{fig:nr}
\end{figure}

Additionally, we study how the number of register qubits $N_R$ required to resolve the ground state scales with the LABS problem size $N$. In Figure \ref{fig:nr}, we plot the minimal $N_R$ as a function of $N$. The number of register qubits required to resolve an energy gap $\Delta E$ is known to scale as 
\be N_R\propto \log_2 \left(\frac{1}{\Delta E}\right)~, \ee 
where in the regime of $N<100$, we find that the energy gap vanishes as $\Delta E\propto N^{-0.323}$.
In our calculations, we use the \texttt{git-labs} solution archive \cite{OPUS2-git_labs-Boskovic} for the LABS solutions up to $N=101$. It is clear that in the regime of up to $N=100$, the resource requirement scales logarithmically with the problem size $N$. 

\section{Conclusion}\label{sec:conclusion} 


In this study, we proposed a QPE-based ground state energy estimation algorithm, QPE-QITE, designed for near-term fault-tolerant quantum hardware. Unlike traditional variational methods, QPE-QITE leverages an ancillary qubit and imaginary time evolution to probabilistically project a quantum state into low-energy eigenstates, eliminating the need for parameter optimization. We demonstrated the effectiveness of the approach using the LABS combinatorial optimization problem and provided resource estimates in the Clifford+T gate basis, highlighting the feasibility of implementation on early fault-tolerant devices.

Beyond the immediate results, QPE-QITE opens several promising avenues for future research. More efficient decompositions of the controlled-QITE operator could significantly reduce the T-gate overhead. This includes exploring optimized synthesis of controlled exponential operators, or leveraging hardware-native gate sets in place of generic Clifford+T-gate decomposition. Additionally, approximate or truncated QITE strategies could improve resource efficiency without sacrificing convergence in practice, particularly when combined with adaptive techniques that modulate $\tau$ during runtime.

QPE-QITE could be generalized to support broader classes of cost functions and Hamiltonians, including those relevant for quantum chemistry, lattice gauge theories, and spin liquids. This may involve integrating sparse Hamiltonian simulation or more advanced quantum signal processing methods to improve precision and scalability. Furthermore, robust error mitigation strategies—such as symmetry verification, post-selection, or probabilistic error cancellation—will be critical to extend QPE-QITE to larger systems under realistic noise models.

A particularly exciting direction is the development of iterative QPE-QITE protocols \cite{Dob_ek_2007}, which can reduce quantum depth and allow partial results to be extracted with fewer resources. These iterative methods are well-suited to current hardware and could serve as transitional algorithms bridging the gap between noisy and fault-tolerant regimes. Coupled with hybrid quantum-classical control loops for estimating and refining $\tau$ or $\alpha$ (the Hamiltonian offset), such protocols may provide practical advantage even before full error correction becomes available.

Benchmarking QPE-QITE on other benchmark problems—such as spin glasses, MaxCut, or quantum chemistry Hamiltonians—would help assess its generalizability and guide hardware-specific optimizations. Cross-comparisons with existing algorithms like VQE, adiabatic methods, and recent quantum eigenvalue transformation techniques could help position QPE-QITE within the broader quantum algorithm landscape.

QPE-QITE demonstrates a compelling alternative to variational quantum eigensolvers for ground state preparation on fault-tolerant quantum computers. Its non-variational nature, performance guarantees, and amenability to scaling make it a strong candidate for early demonstrations of quantum advantage in Hamiltonian optimization. 

\section*{Acknowledgment}
This material is based upon work supported by the U.S. Department of Energy, Office of Science, Office of Science, Advanced Scientific Computing Research (ASCR) program, under Award DE-SC0024325. We also
acknowledge support by the U.S. National Science Foundation under
Award DGE-2152168. A portion of the computation for this work was performed on the University of Tennessee Infrastructure for Scientific Applications and Advanced Computing (ISAAC) computational resources.


\end{document}